\documentclass[12pt,preprint]{aastex}
\shorttitle{A sub-AU outwardly truncated accretion disk}
\shortauthors{McClure et al.}

\begin{document}

\title{A sub-AU outwardly truncated accretion disk around a classical T Tauri star}


\author{M. K. McClure\altaffilmark{1}, W. J. Forrest\altaffilmark{1},
B. A. Sargent\altaffilmark{1}, Dan M. Watson\altaffilmark{1},
E. Furlan\altaffilmark{2}, P. Manoj\altaffilmark{1}, 
K. L. Luhman\altaffilmark{3}, N. Calvet\altaffilmark{4}, C. Espaillat\altaffilmark{4}, P. D'Alessio\altaffilmark{5}, L. W. Hartmann\altaffilmark{4}, 
C. Tayrien\altaffilmark{1}, S. T. Harrold\altaffilmark{1}}

\altaffiltext{1}{Department of Physics and Astronomy, University of Rochester, 
Rochester, NY 14627; melisma@astro.pas.rochester.edu, forrest@pas.rochester.edu, 
bsargent@pas.rochester.edu, dmw@pas.rochester.edu, manoj@pas.rochester.edu, ctayrien@mail.rochester.edu, 
sharrold@mail.rochester.edu}
\altaffiltext{2}{NASA Astrobiology Institute, and Department of Physics and 
Astronomy, UCLA, 430 Portola Plaza, Los Angeles, CA 90095, NASA Postdoctoral Program Fellow; furlan@astro.ucla.edu}
\altaffiltext{3}{Department of Astronomy and Astrophysics, The Pennsylvania State University, University Park, PA 16802; kluhman@astro.psu.edu}
\altaffiltext{4}{Department of Astronomy, The University of Michigan, 
500 Church St., 830 Dennison Bldg., Ann Arbor, MI 48109; ncalvet@umich.edu, ccespa@umich.edu, lhartm@umich.edu}
\altaffiltext{5}{Centro de Radioastronom{\'i}a y Astrof{\'i}sica, Universidad Nacional Aut{\'o}noma de M{\'e}xico, 58089 Morelia, 
Michoac{\'a}n, M{\'e}xico; p.dalessio@astrosmo.unam.mx}

\begin{abstract}
We present the Spitzer Infrared Spectrograph (IRS) spectrum of SR20, a 5--10 AU binary T Tauri system in the 
$\rho$ Ophiuchi star forming region.  The spectrum has features consistent with the presence of a disk; 
however, the continuum slope is steeper than the $\lambda^{-4/3}$ slope of an infinite geometrically thin, optically thick disk, 
indicating that the disk is outwardly truncated.  Comparison with photometry from the literature shows a large increase in the mid-infrared flux from 
1993 to 1996.  We model the spectral energy distribution and IRS spectrum with a wall $+$ optically thick irradiated disk, 
yielding an outer radius of 0.39$_{+0.03}^{-0.01}$ AU, much smaller than predicted by models of binary orbits.  Using a two temperature 
$\chi^2$ minimization model to fit the dust composition of the IRS spectrum, we find the disk has experienced significant grain growth: its
spectrum is well-fit using opacities of grains larger than 1 $\mu$m.  We conclude that the system experienced a significant gravitational perturbation in 
the 1990s.

\end{abstract}


\section{Introduction}

Classical T Tauri stars are typically surrounded by optically thick accretion disks, indicated by strong and broad 
H$\alpha$ emission lines and by a characteristic infrared excess.  Over the last twenty years, observations 
have revealed that many of these young stars are components in binary, and sometimes higher order, systems \citep[e.g.][]{simon92, ghez93}.  
A binary system may produce up to three distinct disks: a circumprimary, circumsecondary, and circumbinary disk 
\citep[][(AL94; JM97)]{artymowicz94, jensen97}.  Gravitational interactions between binaries affect the structure of their 
disks; in particular, the maximum outer radius for a circumstellar disk in a binary system is approximately 0.18 to 0.4 times the 
semi-major axis (AL94), depending on the mass ratio and orbital eccentricity.

SR20 is one such close binary in the $\rho$ Ophiuchi star forming region.  Its components have an angular separation 
ranging from 0.038 to 0.071\arcsec \citep{ghez93, ghez95} which, at a distance to Ophiuchus of 140 pc \citep{dz99} 
corresponds to a projected linear separation of 5.3--9.9 AU.  The secondary is 2.2 magnitudes fainter than the primary at 
2.2 $\mu$m \citep{ghez93} and therefore is less massive than the primary, diskless, or both.  H$\alpha$ emission has been detected from SR20 with 
equivalent widths ranging from 15 to 21 \AA \citep[][(W05)]{r80, ba92, wilking05}.  Further observations by \citet{tbc03} 
did not find a spectro-astrometric signature in the H$\alpha$ line, indicating that there is little to no accretion in the secondary relative to the primary.  
Based on the lack of evidence for a circumsecondary disk, we assume that the infrared excess of SR20 is dominated by a circumprimary accretion disk.

Here we present Spitzer Space Telescope Infrared Spectrograph (IRS) \citep{houck04} observations of the SR20 system.  
We model the structure of the system and the dust it contains and analyze the nature of the SR20 system.  
While instances of sculpture of proto-planetary accretion disks by planetary 
or stellar companions \emph{within} the disk are becoming widely known \citep[e.g. JM97,][]{furlan07}, SR20 is an apparently 
rarer example of sculpture from \emph{without}.

\section{Observations and Data Reduction}

We observed SR20 on 2006 April 16 (Spitzer AOR 12698368), with the short wavelength (SL; 5.2--14 $\mu$m) and long wavelength (LL; 
14.0--36.1 $\mu$m) low spectral resolution ($\lambda$/$\Delta\lambda$=60--120) IRS modules.  At each order of each module, the object 
was observed twice, once in each of the nominal nod positions, $1/3$ of the way from the ends of the slit, with exposure times for each 
observation of 6 seconds in SL and 14 seconds in LL.  The spectrum was extracted from the Spitzer Science Center (SSC) S14.0 pipeline basic calibrated data 
using the SMART software tool \citep{higdon04} following the same procedure as \citet{furlan06} with the exception of background 
subtraction, which was done using the opposite nod position.

Assuming nod-to-nod flux differences were caused by slight mispointing, for each module we scaled the nod with the lower flux to 
match the flux level of the nod with the higher flux.  Scale factors were typically within a few \% of unity.  
The resulting nods were averaged to obtain the final spectrum, and the spectral uncertainties are estimated to be half 
the difference between the two independent spectra from each nod position.  We estimate the spectrophotometric accuracy of 
the final result to be around 5\%.

\section{Analysis}
\subsection{Spectral Energy Distribution}
\label{SED_section}

\begin{figure}
%\epsscale{0.7}
\plotone{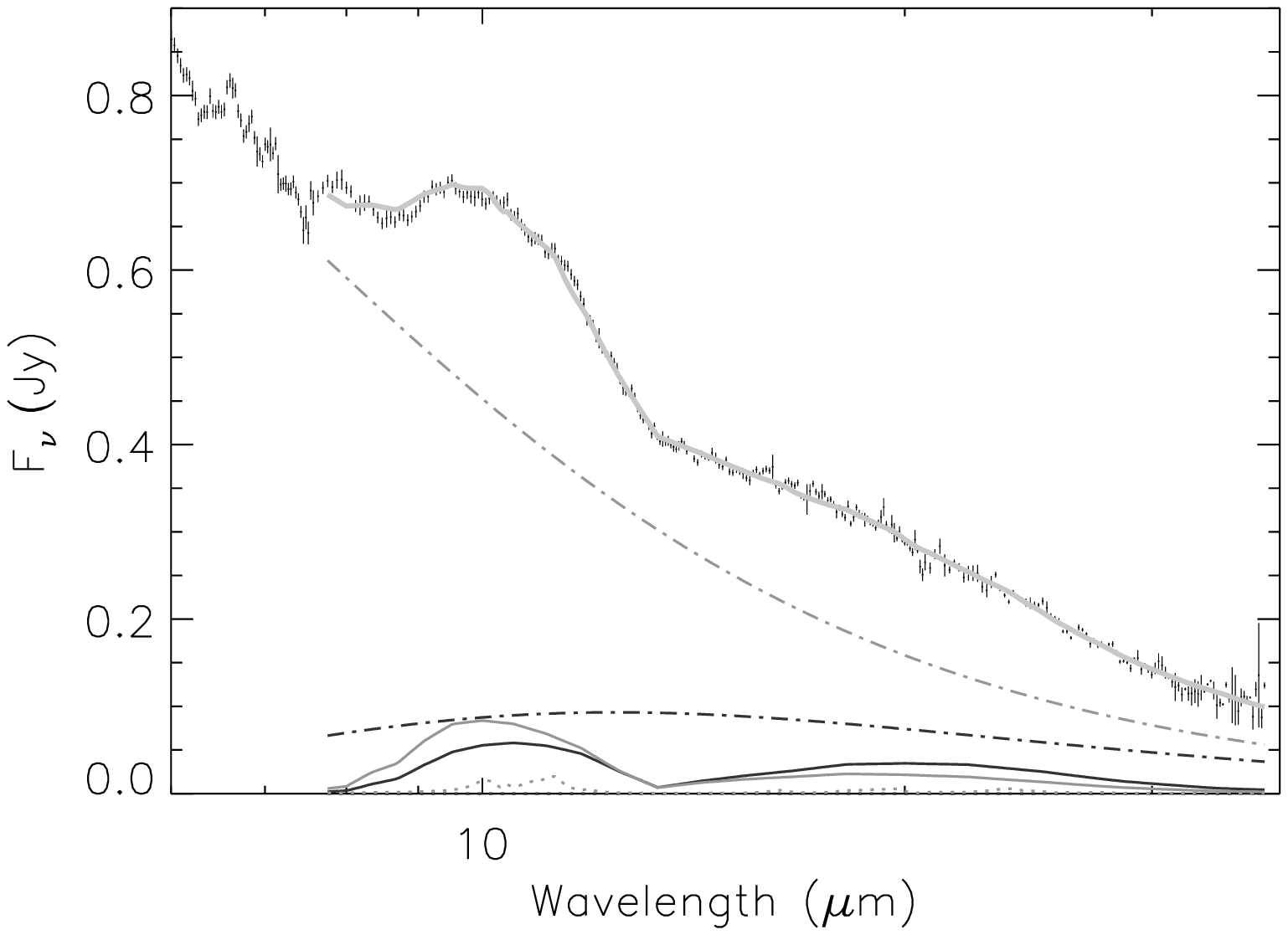}
\caption{Dust composition model overplotted on the extinction corrected IRS spectrum.  Light grey and dark grey represent 410 K and 
1221 K components respectively.  Components are black bodies (dash/dot), large amorphous olivine (solid), and forsterite (dotted).  
Forsterite is present at less than $1\sigma$ strength and not considered significant. 
\label{SR20_dust}}
\end{figure}

The 5--36 $\mu$m IRS spectrum of SR20 (Figure \ref{SR20_dust}) has a clear infrared excess and weak silicate dust emission 
at 9.8 and 18.0 $\mu$m, consistent with the presence of a circumstellar disk.  This excess cannot be attributed 
to an optically thin shell, as in mass-losing giant stars, since the 5--8 $\mu$m H$_2$O absorption features indicate that 
the underlying continuum is optically thick. 

Here we adopt a spectral type of G7 and $A_V$ = 5.7 mag, determined from optical data by W05.  
To correct the photometry and IRS spectrum for continuum slope fitting, we applied a ``standard'' extinction correction model with 
$A_V/\tau_{9.7}$ = 12.5 and $R_V$ = 3.1 \citep[][D03]{draine03}.  For the dust composition fitting, we corrected more carefully over the silicate features 
by scaling our previous correction model to $A_V/\tau_{9.7}$ = 25, which may be more appropriate for silicate features in dark clouds \citep{chiar07}.  
For consistency, we compared results for both dust composition and disk structure and they are independent of the correction used. 

\begin{figure}
%\epsscale{0.7}
\plotone{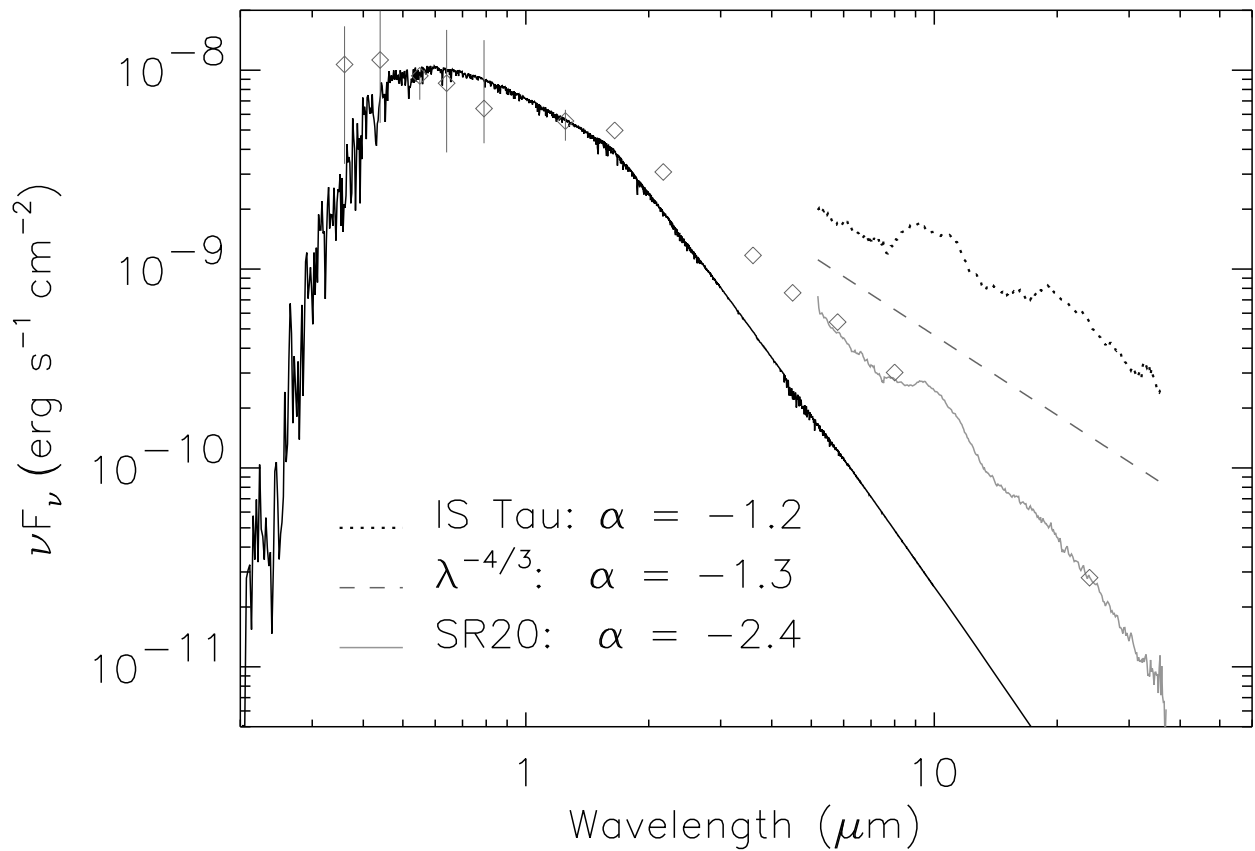}
\caption{``Current'' SED of SR20, with 1983-84 UBVRI \citep{ba92}, 1999 JHK (2MASS), 2004 IRAC (C2D), and MIPS 24 $\mu$m photometry 
with our 2006 IRS spectrum.  Data were extinction corrected using $A_V$=5.7 and the D03 law.  UBVRI data are plotted with bars to indicate the extent of 
variability.  Overplotted are the IRS spectrum of IS Tau ($\times$ 15) and a line representing $\lambda^{-4/3}$. \label{SR20_SED}}
\end{figure}

\begin{figure}
%\epsscale{0.7}
\plotone{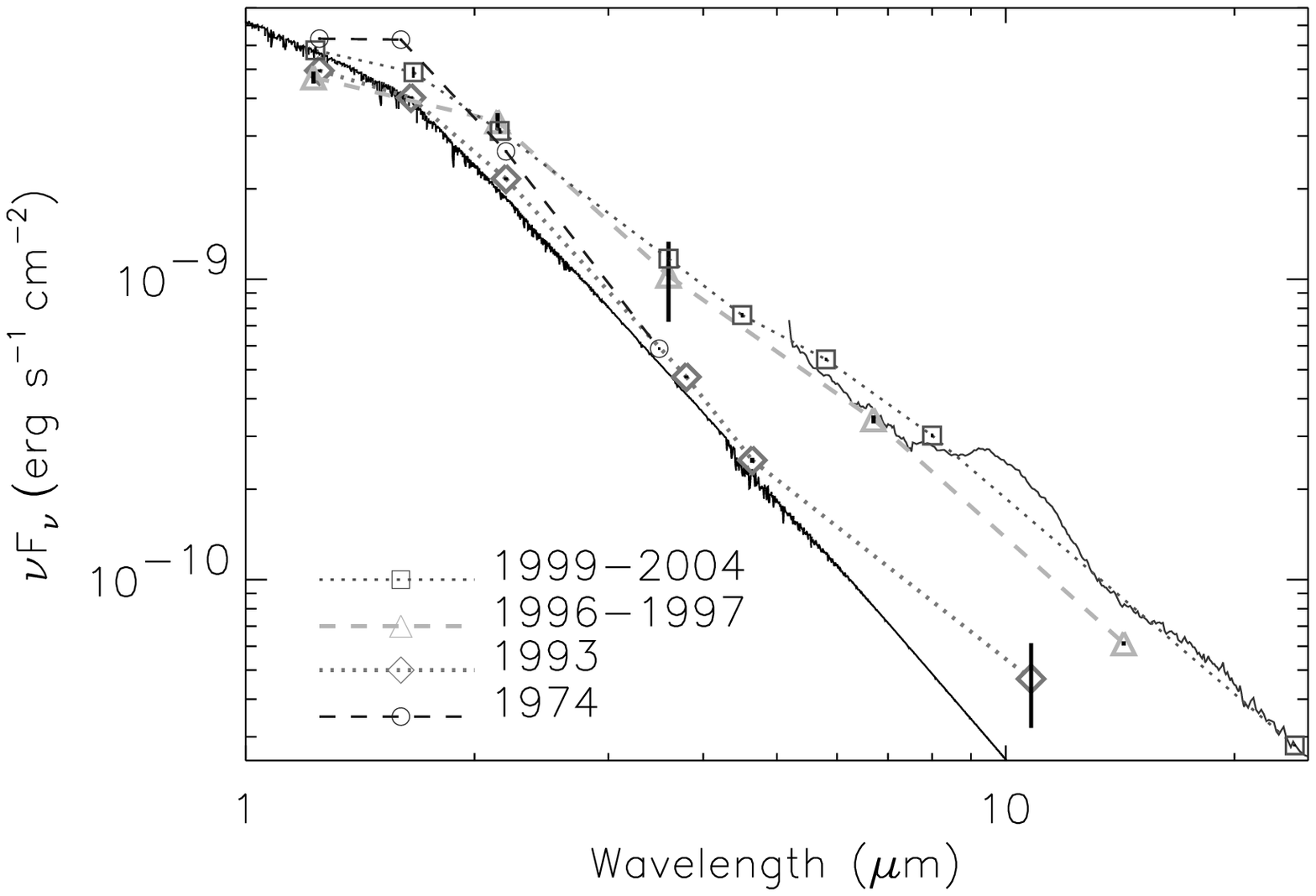}
\caption{SR20 photometry from four epochs: 1974: JHKL \citep{rss76}; 1993: JHKL'M'N (JM97); 1996-7: JK (DENIS), 3.6 $\mu$m 
(ISO-PHT, sky+source), 6.7 and 14.3 $\mu$m \citep{bontemps01}; 1999-2004: JHK (2MASS), IRAC (C2D), 24 $\mu$m 
(Extracted from MIPS AOR 4321280 with an aperture of radius 14.94\arcsec\, sky annulus from 29.88 to 42.33\arcsec\, and aperture 
correction factor of 1.143 \citep{su06} using the SSC's APEX software package \citep{mm05}).  Photometry is plotted with 
error bars, most of which are smaller than the plot symbols. \label{SR20_3epoch}}
\end{figure}

After exinction correction, we constructed two SEDs using the $A_V/\tau_{9.7}$ = 12.5 corrected spectrum. 
The first contains optical data, the near infrared data closest in time to our spectrum, the IRS spectrum and 
a 5600 K, solar metallicity NextGen stellar photosphere model \citep{nextgen} (Figure \ref{SR20_SED}).  We chose to normalize 
the photosphere model to the 2MASS J-band flux density rather than that at I-band due to the greater variability at I (1.0 magnitude vs. 0.4 magnitudes at J).  
It is unlikely that SR20 suffers from significant veiling at J, since the line strengths in the \citet{lr99} K-band spectrum are comparable with a late G photosphere.
The second SED shows the near infrared data, sorted into four epoch ``bins'': 1974, 1993, 
1996-1997, and 1999-2004 (Figure \ref{SR20_3epoch}).  Significantly, in 1974 and 1993 the excess from 2 to 10 $\mu$m 
was small but increased substantially by 1996.  Changes in the 2--10 $\mu$m excess since then have been small.

It can be seen from Figure 2 that SR20 exhibits excess emission above the photosphere over $\lambda$ = 1.25--36 $\mu$m and that the slope of the 
continuum emission is much steeper than $\lambda^{-4/3}$, the slope of an infinite geometrically-thin, optically-thick irradiated disk 
\citep{hartmann99}.  There is a correlation between the continuum slope and dust settling in the disk; for the IRS Taurus sample, 
the bluest and presumably most settled T Tauri disks had continuum slopes approaching, but not equal to, $\lambda^{-4/3}$ 
\citep{furlan06}, as demonstrated by the IRS spectrum of IS Tau, also in Figure \ref{SR20_SED}.  We interpret a slope steeper 
than $\lambda^{-4/3}$ to indicate that the circumstellar disk is outwardly truncated.

\subsection{Wall + Disk Model}
\label{Sys_Model_section}

To investigate whether outward truncation is plausible and determine the degree to which it occurs, we modeled the disk over a range of radii.  We began by modeling 
the inner edge of the disk with a vertical ``wall'' located at the dust destruction radius \citep{natta01}, calculated following the methods of \citet{dalessio05}, 
which include the wall atmosphere.  The wall emission depends on its radius, $R_{in}$, height $Z$, and the inclination to the 
line of sight, $i$.  In turn, $R_{in}$ depends on the dust composition and size distribution in the wall and on the sum of the stellar, $L_*$, and accretion, 
$L_{acc}$, luminosities.  We use a size distribution  $\propto a^{-3.5}$ between $a_{\rm min} = 0.005 \mu$m and $a_{\rm max}$.  For $L_*$, we adjusted the value from 
W05 to 140 pc, yielding 6.76 L$_{\odot}$.  For a list of stellar parameters, see Table \ref{Stellar parameters}.\citet{natta06} estimated an upper limit to 
$L_{acc}$ of 0.03 L$_{\odot}$, using the luminosity of the Pa$\beta$ line. They adopted an extinction of $A_J$ = 1.3, which corresponds to $A_V = 4.6$, lower than the 
one used here. Scaling up the J band flux according to the difference of extinction values, we get an upper limit to the accretion luminosity of 0.05 L$_{\odot}$, which is 
much lower than $L_*$.  Therefore, we neglect the accretion luminosity in the calculation of the dust destruction radius and use as input parameters the maximum grain size, 
$a_{max}$, the inclination, and the height of the wall.  Varying $a_{\rm max}$  changed $R_{in}$ and the peak-to-continuum ratio of the silicate features, which formed 
in the wall atmosphere, with large $a_{\rm max}$ producing a wall radius closer to the star and a lower peak-to-continuum ratio.  We were able to fit most of the SED with 
just the emission from a wall with spherical solid amorphous olivine grains that have grown to 1.0 $\mu$m.  With these grains, the dust destruction radius is located at 
0.34 AU. In the vertical wall approximation, the maximum emission occurs for $\sim 50 \degr$ \citep{dull01}. At this inclination, best fit is for a wall height $Z$ = 
2.11 R$_*$, or 2.6 scale heights, which is lower than the expected range of 4-5 scale heights \citep{dull01}. Lower inclinations would require higher walls to fit 
the near-infrared excess.  The wall emission alone is insufficient to explain the IRS spectrum past 10.0 $\mu$m, so we added a very small flat, optically thick 
irradiated disk component \citep{hartmann99}.  The temperature distribution of the disk component is given by

\begin{figure}
%\epsscale{0.7}
\plotone{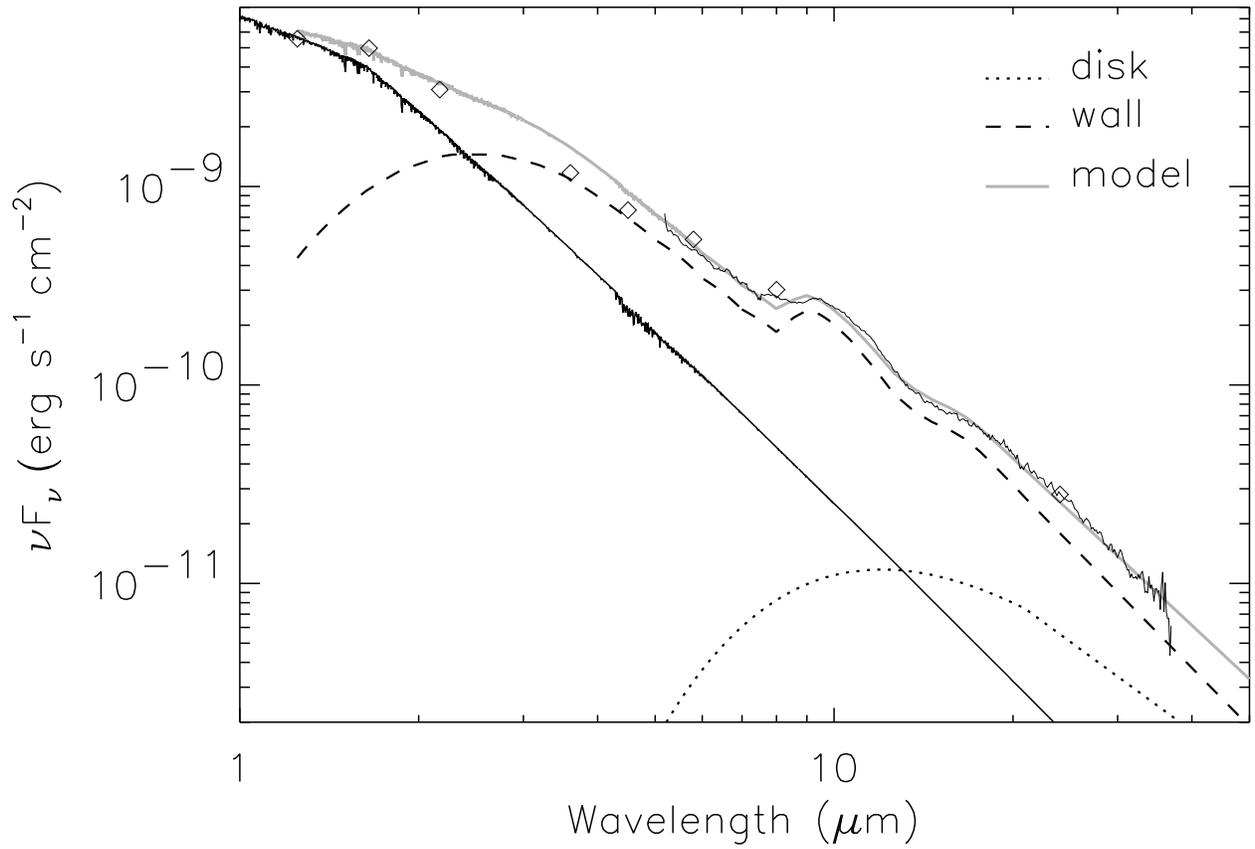}
\caption{Best fitting model (grey solid line):  d = 140 pc, $i$ = 50\degr\, $R_{\rm in}$ = 0.34 AU, and $R_{\rm out}$ = 0.39 AU.
\label{Radius_grid}}
\end{figure}

%\begin{equation}
$T(r) = \left[\frac{I_0}{\sigma}\left[{\rm sin}^{-1}\frac{R_{*}}{r} - 
\frac{R_{*}}{r}\left(1 - \frac{{R_{*}}^2}{r^2}\right)^{\frac{1}{2}}\right]\right]^{\frac{1}{4}}$
%\end{equation}

\noindent where $r$ is the cylindrical radius, and $I_0=\frac{\sigma{T_{e}}^4}{\pi}$.  From 1.25--971 $\mu$m, we numerically integrated the corresponding flux density 
over the range of radii from $R_{\rm in}$, the wall radius, to $R_{\rm out}$, the outer radius.  By adding a disk component with $R_{\rm in}$ = 0.34 AU, $R_{\rm out}$ 
= 0.39$_{+0.03}^{-0.01}$ AU, and $i$ = 50$\degr$ to the wall model at the same inclination with 1.0 $\mu$m solid olivine grains, the SED of SR20 is well fit (Figure 
\ref{Radius_grid}).  The only discrepancy is on the longer wavelength side of the silicate features, which could be fit better with slightly larger grains.  
As noted, we could get similar fits in the near-IR with lower inclinations and taller walls. However, the emission of the optically thick disk scales as 
$ {\rm cos} i$, increasing as the inclination decreases. As a result, to obtain the same fit to the mid-infrared, the disk radius would have to be smaller.  
A similar conclusion could be expected if the wall were round and its maximum emission occurred at low inclinations \citep{IN05}. Therefore, 0.39 AU
is an upper limit for the radius of the disk. 

To understand the relationship between the binary companion and circumprimary disk, we need information about the binary orbit.  
We were unable to determine a unique solution (there are only three separation measurements in the literature), but we placed rough 
constraints on the semi-major axis and eccentricity.  In the limit of the eccentricity approaching unity, the semi-major axis is half the maximum separation. 
Therefore, regardless of the eccentricity, the semi-major axis of SR20 must be greater than 5 AU.  For our preliminary orbit estimate, values between 8 and 
12 AU fit best along with lower eccentricities of 0.2--0.3.

\subsection{Dust Composition Model}
\label{Dust Comp}

Although our wall $+$ irradiated disk model produced a good fit to the IRS spectrum, it uses only one silicate component at a time with a fixed set of $a_{\rm max}$.  
To perform a more detailed analysis, we calculate a two temperature model of the silicate dust composition by $\chi^2$ minimization with respect to weights proportional 
to dust mass, black body solid angle, and temperature for black bodies and dust grains \citep{sargent06, kastner06, chen06} 
with optical constants from \citet{day79} and \citet{koike03}.  Grains larger than 8 $\mu$m in radius, which do not contribute 
to the silicate features, could still contribute to the black body components.  The best fit ($\chi_\nu$ = 4.0) occurs at 
410 and 1221 K with the components listed in Table \ref{Stellar parameters}, which are shown in Figure \ref{SR20_dust}.  
Interestingly, the only significant components besides the black bodies are large (5 $\mu$m) porous (60\% vacuum), amorphous olivine 
(Mg$_2$SiO$_4$) grains, which is consistent with the 10 $\mu$m peak to continuum ratio of 1.35:1.  Submicron amporphous olivine and pyroxene, 
crystalline silica, and large amorphous pyroxene, all of which are typical of T Tauri disks, were included in the modeling but found not to be present.  
Forsterite was present in the best-fit model, but at less than 1$\sigma$ levels, so we do not consider it to be significant.  For $d$ = 140 pc, our fit yields 9.55 $\times$ 
$10^{-5}$ $M_{\rm lunar}$ of large amorphous olivine grains, combining the contributions at both temperatures, with an upper limit of 1.36 $\times$ $10^{-5}$ $M_{\rm lunar}$ 
on sub-micron grains.

\section{Discussion and Conclusions}
\label{discussion}

The SED, system geometry, and dust composition shed light on the nature of the SR20 system, while simultaneously 
raising further questions.

The IRS spectrum is well-fit by our composite wall $+$ circumprimary disk truncated beyond 0.39 AU.  Outward truncation could be caused by 
the orbit of a companion or lack of replenishment from a circumbinary disk.  Using the tidal truncation models of AL94, the range of likely eccentricities (0.2-0.3) 
and semi-major axes (8-12 AU) of the A-B orbit are not extreme enough to cause truncation.  Even with the lower limit on the semi-major axis, 5 AU, and the highest 
order resonances and eccentricities tested in AL94, the minimum stable outer truncation radius caused by the known binary is 0.9 AU, considerably larger than the 0.39 AU 
we derive (AL94).  However, the AL94 models may not be appropriate for a disk with such a small ratio between the outer and inner radii, so we cannot {\it absolutely} 
rule out truncation by the known binary, at least on this basis.  Alternatively, it may be that SR20 is a hierarchical triple.  A companion object with an eccentricity 
$<$ 0.8 could orbit in the region between 0.98 AU and 2.17 AU from the primary, causing a disk truncation at 0.39 AU.  Such a close, low mass companion would be difficult 
to detect.

Based on the 0.39 AU truncation radius, we speculate that replenishment of the circumprimary disk by a circumbinary disk occurred in the past.  
A disk between 0.1 and 10 AU would accrete on a timescale of 1,000--100,000 years \citep{quillen04} and the median age of the Ophiuchus cluster is 2.1 Myr (W05).  
SR20 has been observed at 800, 850, and 1300 $\mu$m with only 3 $\sigma$ upper limit results, implying less than $5.4$ $\times$ $10^{4}$ $M_{lunar}$ (540 $M_{lunar}$ 
of dust) in disk material \citep{aw07}.  Low circumbinary disk masses are typical of binaries with semi-major axes between 1 and 50 AU \citep{jensen94}, 
so submillimeter non-detection does not indicate a depleted circumbinary disk; however, it is unlikely that lack of replenishment is responsible for the 0.39 AU 
truncation radius.

A nearby companion truncating the circumprimary disk is perhaps also the best explanation for the photometric variability.  Before 1993, the infrared excess in 
SR20 started in the N band, indicating that there were few small dust grains close to the central star.  
However, the H$\alpha$ emission indicates that a gaseous accretion disk, perhaps mixed with much larger planetesimals, was present.  
Between 1993 and 1996, the infrared excess increased sharply, indicating an increase in the amount of dust in the disk, which could 
be explained by a gravitational interaction between the known binary and a third, unseen, body, causing the unseen companion to orbit 
closer to the primary, perturbing the circumprimary disk.  Coupling between the gas and dust could confine the newly created dust to the accretion disk.  
The appeareance of the observed excess over 3 years, preceded by 20 years without an excess and followed by 10 years of little variation in the observed excess, seems 
consistent with truncation by a nearby source with a shorter period than is plausible for the known companion. 

The silicate dust composition of the system is not typical of a Class II object; our fit indicates that the optically thin 
portion of the disk is comprised primarily of large grains ($>$ 1 $\mu$m).  Although having a large fraction of large grains is consistent 
with debris disks or advanced dust processing in the inner regions of a protoplanetary disk \citep{chen06}, this is the first time we have seen a Class II disk 
with negligible submicron grains.  Taken with the aforementioned explanations of the 1993--1996 variability, the large grains 
could support a collision scenario:  gravitional interactions between the known secondary and the unseen companion cause collisions 
between planetesimals, producing dust grains that are detected as the current disk. If there were a population of planetesimals -- invisible in our spectra -- comprising 
an extension of the disk beyond the outer truncation radius of the dust, truncation by the known companion would be more probable.

Near infrared interferometry and orbital monitoring would be necessary to determine the nature of this complex system more precisely.

\acknowledgements
This work is based on observations made with the {\it Spitzer Space Telescope}, 
which is operated by the Jet Propulsion Laboratory, California Institute of Technology, 
under NASA contract 1407 and made use of the ADS, SIMBAD, and Vizier utilities. Support was provided by NASA through 
contract number 1257184 issued by JPL/Caltech, JPL contract 960803 to Cornell University, 
and Cornell subcontracts 31419-5714 to the University of Rochester.  

\begin{deluxetable}{lccc}
\tabletypesize{\scriptsize}   %\small (11pt), \footnotesize (10 pt), \scriptsize (8pt)
\tablewidth{\linewidth}
\tablecaption{Stellar and model properties \label{Stellar parameters}}
\tablehead{
\colhead{Component} & \colhead{Value} & \colhead{Reference}
}
\startdata
\multicolumn{4}{c}{Stellar parameters}\\
\hline
$SpT$ & G7 & W05 \\
$T_{e}$ (K) & 5584  & W05 \\
$L_{*}$ (L$_{\odot}$) & 6.76 & W05, adjusted$^a$ \\
$M_{*}$ (M$_{\odot}$) & 1.86  & \citet{siess2000}\\
$R_{*}$ (R$_{\odot}$) & 2.57 & \citet{siess2000} \\
$L_{acc}$ (L$_{\odot}$ & $<$ 0.05 & \citet{natta06}, adjusted$^a$ \\
$A_V$ (mag) &  5.7 & W05 \\
\hline
\multicolumn{4}{c}{Dust Components $>$ 1$\sigma$}\\
\hline
Parameter & Value$^b$ & $1\sigma$ uncertainty$^c$ \\
\hline
(410 K) \\
black body & $7.12 \times 10^{-17}$ &  $3.84 \times 10^{-18}$ \\
large Mg$_2$SiO$_4$$^d$ & $3.41 \times 10^{-20}$ & $5.18 \times 10^{-21}$ \\
\hline
(1221 K) \\
\hline
black body &  $2.56 \times 10^{-17}$ &  $4.70 \times 10^{-19}$ \\
large Mg$_2$SiO$_4$$^d$ & $3.58 \times 10^{-21}$ & $5.97 \times 10^{-22}$ \\
\tableline
\enddata
\tablecomments{
$^a$ See section 3.2 for details.
$^b$ Blackbody components are solid angles in units of sterradians, while dust components are column density times solid angle in g/cm$^2$.  
$^c$ Uncertainties are the increments away from the best fit values enough to increase $\chi_\nu$ by 1.0.  $^d$ Amorphous olivine}
\end{deluxetable}

%\clearpage
% Figures and tables here

\end{document}